# Surface Van Hove Singularity Enabled Efficient Catalysis: The Cases of CO Oxidation and Hydrogen Evolution Reactions

Liangliang Liu[1,3], Chunyan Wang[1,2], Liying Zhang[1], Chengyan Liu[1], Chunyao Niu[2], Zaiping Zeng[1], Dongwei Ma[1,3*], and Yu Jia[1,2,3*]

[1] Key Laboratory for Special Functional Materials of Ministry of Education, Collaborative Innovation Center of Nano Functional Materials and Applications, School of Materials Science and Engineering, Henan University, Kaifeng 475004, China

[2] International Laboratory for Quantum Functional Materials of Henan, and School of Physics and Microelectronics, Zhengzhou University, Zhengzhou 450001, China

[3] Key Laboratory for Quantum Materials of Henan, and Center for Topological Functional Materials, Henan University, Kaifeng 475004, China

**Corresponding Author**:

madw@henu.edu.cn (Dongwei Ma), jiayu@henu.edu.cn (Yu Jia)

**Abstract:** Surface Van Hove singularity (SVHS), defined as the surface states near the Fermi level ($E_F$) in low-dimensional systems, triggers exciting physical phenomena distinct from bulk. We herein explore theoretically the potential role of SVHS in catalysis taking CO oxidation reaction as prototype over graphene/$Ca_2N$ (Gra/$Ca_2N$) heterojunction and $Pt_2HgSe_3$ (001) surface. It is demonstrated that both systems with SVHS could serve as an electron bath to promote $O_2$ adsorption and subsequent CO oxidation with low energy barriers of 0.2 ~ 0.6 eV for Gra/$Ca_2N$ and $Pt_2HgSe_3$ (001) surface. Importantly, the catalytically active sites associated with SVHS are well-defined and uniformly distributed over the whole surface plane, which is superior to the commonly adopted defect or doping strategy, and further the chemical reactivity of SVHS also can be tuned easily via adjusting its position with respect to $E_F$. Our study demonstrates the enabling power of SVHS, and provides novel physical insights into the promising potential role of VHS in designing high-efficiency catalysts.

*oxidation; Density functional theory*

Van Hove singularity (VHS) is a saddle point in the electronic band structure where the density of states (DOS) diverges logarithmically, exhibiting the most characteristic feature of existing highly localized electronic states [1-3]. The electronic instability in the vicinity of VHS promotes diverse novel phases of matter, such as superconductivity [4-5], ferromagnetism [6] and charge density wave [7], which makes the study of VHS physics the research frontier of condensed matter physics and materials science. If the VHS exists in reduced dimensional systems, such as surface systems or two-dimensional (2D) materials, and serve as the surface states near the Fermi level ($E_F$), we name it as surface Van Hove singularity (SVHS). In stark contrast to the VHS in bulk which is remote from the Fermi level ($E_F$) and insensitive to external perturbations, SVHS can be tuned to be around or even well-aligned with $E_F$, delivering rich surface physical and chemical properties waiting for exploration. This can be achieved in a much easier way via surface engineering, such as chemical doping [8-9], intercalation [10-11], constructing heterojunction [12-13], gating [14] or twisting [15-16], thanks to the emergence of novel surface or 2D layered materials with rich characteristics and functionalities.

Various surface or 2D material systems with SVHS have been realized, and intriguing surface or interface phenomena has been subsequently reported, including interface superconductivity, surface charge density wave, and optical adsorption [17-23]. Monolayer gaphene on SiC substrate or bilayer boron nitride via alkali-metal intercalation enables the appearance of SVHS, which induces the phonon-mediated superconductivity through the strong electron-phonon coupling at the vicinity of SVHS [17-18]. Optical absorption has also been found to be significantly enhanced in twisted bilayer graphene with SVHS locating at the $E_F$, and G band resonance has been observed [19-21]. Due to the charge transfer from $Ca_2N$ electride, the $E_F$ is pushed up to be well-aligned with the π*-band VHS of graphene, and hence SVHS with large DOS appears in the graphene/$Ca_2N$ electride (Gra/$Ca_2N$) heterojunction, as theoretically predicted [24]. The efficient charge transfer nature has been confirmed by later experiment on heterojunctions formed with multilayered graphene on $Ca_2N$ [25]. Besides the aforementioned tuning strategies, the recently discovered topological materials with intrinsic surface states close to $E_F$ could natively serve as SVHS, which originates from the nontrivial topological nodal line of the bulk, protected with certain symmetry. A prototypical example is $Pt_2HgSe_3$ film with cleaved (001) surface [26-27].

One the peculiar characteristics of the SVHS is the highly localized electronic states [2-3] that could benefit the charge exchange between them, and thus holds promising potential for efficient catalysis [28-29]. In this contribution, we examine the SVHS catalysis choosing prototypical chemical reactions on ground of the aforementioned two well-established platforms of Gra/$Ca_2N$ heterojunction [24] and $Pt_2HgSe_3$ film with (001) surface [26-27]. All first-principles calculations are performed within the framework of density functional theory using the Vienna *ab initio* simulation package with the projector-augmented wave method [30-32] (see Computational details in Supporting Information). The reaction barrier is computed using the climbing image nudged elastic band method [33]. We demonstrate the enabling power of SVHS on CO oxidation and hydrogen evolution reactions. The SVHS in both systems is found to serve as electron bath to promote $O_2$ adsorption and subsequent CO oxidation with low kinetic energy barriers of 0.37 eV for Gra/$Ca_2N$ and 0.59 eV for $Pt_2HgSe_3$. Two features of SVHS catalysis have been further identified. One is that the active sites originating from the SVHS are well-defined and uniformly distributed over the whole surface plane. The other is the tunability of the chemical activity driven by SVHS through adjusting its relative energy position with respect to $E_F$ as for Gra/Ca$_2$N heterojunction. Our study drives Van Hove singularity to the catalytic applications, and provides promising degree of freedom towards the search and design of catalysts.

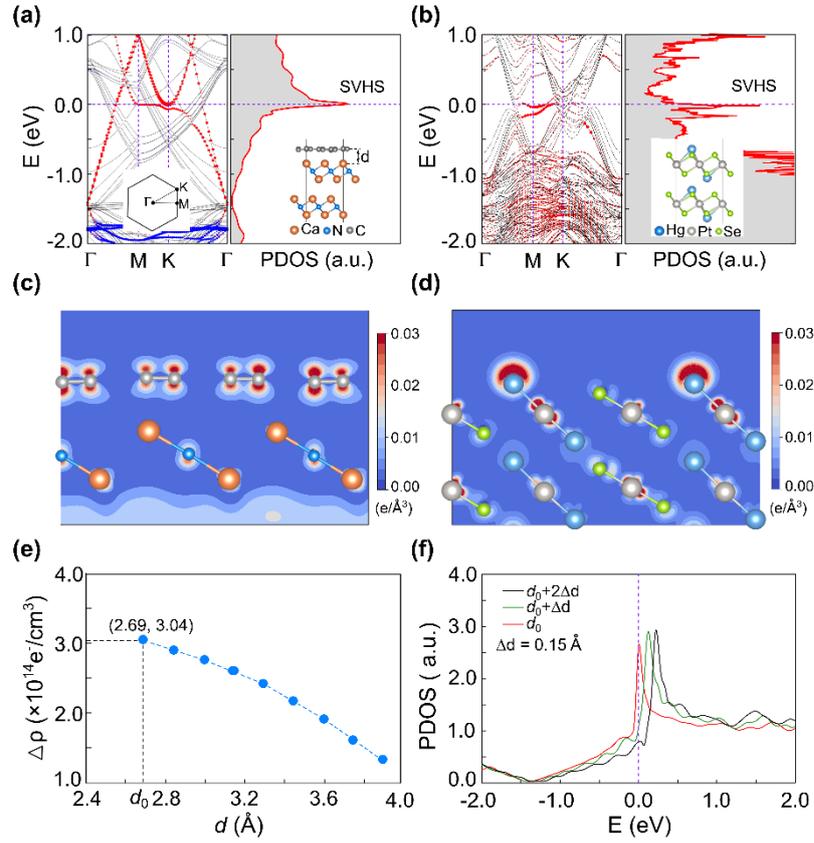

Figure 1: (a) Calculated band structure of the Gra/Ca$_2$N and partial DOS (PDOS) for graphene. Here, the bands projected onto graphene and the first Ca$_2$N atomic layer are respectively displayed with red and blue circles. $d$ is the interlayer distance between graphene and the topmost Ca layer. (b) Surface band structure of a ten-layer thick Pt$_2$HgSe$_3$ slab. The bands projected onto the first layer are displayed with red circles. The energy zero represents $E_F$. (c) and (d) Calculated partial charge density around the SVHS (electron energy between -0.2 and 0 eV) for Gra/Ca$_2$N and Pt$_2$HgSe$_3$ surfaces respectively, plotted on the (110) face. (e) Net average charge ($\Delta\rho$) and (f) PDOS for the graphene layer in Gra/Ca$_2$N heterojunction as a function of the interlayer distance $d$. In (e) and (f), $d_0$ is the optimal distance, 2.69 Å.

To begin with, we study the CO oxidation reaction on Gra/Ca$_2$N heterojunction and Pt$_2$HgSe$_3$ (001) surface. The electronic band structures of the proposed systems are given in Figs. 1(a) and (b). In both systems, the SVHS at M-point appears well-aligned with E$_F$. For Gra/Ca$_2$N heterojunction, VHS in the free-standing graphene is about 1.6 eV above the E$_F$ [34]. Thanks to the strong charge transfer nature of electride, E$_F$ is pushed up, and aligns with the SVHS. As demonstrated by the PDOS, SVHS exhibits a unique feature of large DOS. The charge density plot of Fig. 1(c) shows that the charges for the SVHS are highly localized, distributing mainly on two lobes of the 2$p_z$ orbitals of the carbon atoms. For Pt$_2$HgSe$_3$ (001) surface, the surface states are nontrivial with a unique saddle-like energy dispersion, and the associated SVHS is natively positioned near the E$_F$ [26-27]. The presence of SVHS in such a system is also justified by the large DOS near the saddle point [see Fig. 1(b)]. The charges of the SVHS are found to be highly localized on the surface Hg atoms [see Fig. 1(d)]. We further show in Fig. 1(e) the modulation of net charges in graphene as a function of the interlayer spacing $d$ of the Gra/Ca$_2$N heterojunction. Due to the electron donating from the electride layer, graphene component receives a high $n$-doping at the optimal distance $d_0$, with charge density reaching up to $3.04 \times 10^{14}$ e$^-$/cm$^2$. Strikingly, the net charge $\Delta\rho$ decreases sublinearly with respect to the interlayer distance $d$ [see Fig. 1(e)]. We note that considerably large net charge density ($> 1 \times 10^{14}$ e$^-$/cm$^2$) in the graphene layer can be easily reached as long as the interlayer distance is larger than 4 Å. Accordingly, the position of E$_F$ is tuned around the SVHS [see Fig. 1(f)], through which it is expected the chemical and catalytic activity of the graphene layer can be modulated.

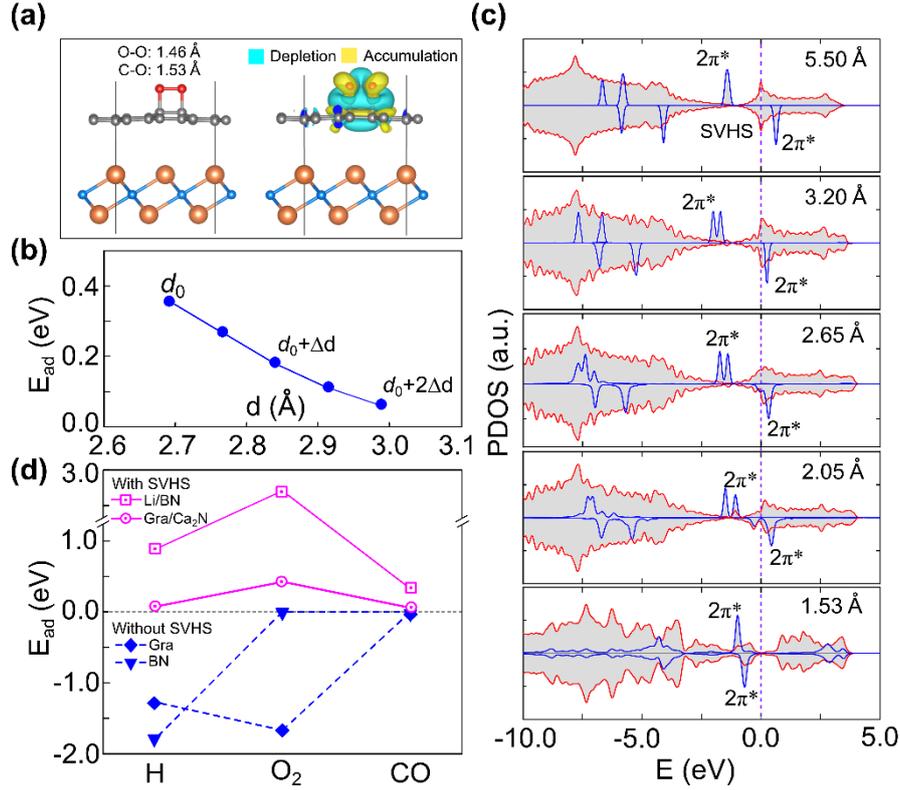

Figure 2: (a) The adsorption configuration and isosurface (0.015 e$^-$/Å$^3$) of charge density difference for an O$_2$ molecule on the Gra/Ca$_2$N surface. (b) The change of O$_2$ adsorption energy as a function of interlayer distance $d$. (c) The PDOS of O$_2$ (blue line) and graphene (red line) during the process of O$_2$ approaching to the Gra/Ca$_2$N. The red sphere in (a) denotes the O atom. In (b), $d_0$ is the optimal distance. (d) The adsorption energies, for H, O$_2$, and CO on Gra/Ca$_2$N, the free-standing graphene, Li/BN and the free-standing BN monolayer.

We next study the CO oxidation reaction on the basal plane of graphene in Gra/Ca$_2$N heterojunction. We firstly evaluate the adsorption of O$_2$. Significantly, O$_2$ molecule is chemisorbed exothermically on Gra/Ca$_2$N heterojunction. It should be noted that its adsorption ability towards O$_2$ is found to be nearly independent on the absorption sites with the adsorption energies ~ 0.3 eV [see Fig. S1 in in Supporting Information]. The most stable configuration is shown in Fig. 2(a), from which we can see the strong chemical bonds between C-O bonds from (1.53 Å), and the O-O bond (1.46 Å) is significantly elongated, compared with the free $O_2$ molecule (1.23 Å). We have further studied the $O_2$ adsorption under the various interlayer distances between graphene and $Ca_2N$, corresponding to the different energy intervals between the SVHS and $E_F$ [see Fig. 1(f)]. With increasing in the interlayer distance, the adsorption energy of $O_2$ is significantly reduced [Fig. 2(b)]. It is suggested that tuning the position of SVHS with respect to $E_F$ by external means could be an efficient strategy to modulate the catalytic behavior of SVHS. Moreover, Bader charge analysis [35] shows that there are 0.83 $e^-$ transferred from the heterojunction to the adsorbed $O_2$, from the charge density difference, which mainly occupy its anti-bonding orbital [see Fig. 2(a)]. The evolution of the electronic states as a function of the distance between $O_2$ and Gra/$Ca_2N$ is shown in Fig. 2(c). Obviously, once the $O_2$ approaching to Gra/$Ca_2N$, the major spin-down component of $O_2$ $2\pi^*$ orbitals are gradually shifted below $E_F$, while the SVHS states shifted up with respect to $E_F$. This clearly indicates that the SVHS serves as an electronic bath to contribute to $O_2$ adsorption through electron transfer.

To consolidate the decisive role of SVHS catalysis on the CO oxidation, we study of the adsorption of H atom, $O_2$, and CO molecules on systems inheriting SVHS, including the aforementioned Gra/$Ca_2N$ and Li intercalated bilayer boron nitride (Li/BN) [see Fig. S2 in in Supporting Information], in comparison with their counterparts without the presence of SVHS, i.e., free-standing graphene and the free-standing BN. The systems with SVHS exhibit much higher adsorption energies than their counterparts without SVHS. Indeed, Li/BN with a SVHS laying 0.15 eV below $E_F$ exhibits a drastic increase in H and $O_2$ adsorption energies by amount of 3.0 eV, compared with the intrinsic BN. In parallel, the binding strengths of H and $O_2$ on Gra/$Ca_2N$ are enhanced by ~ 1.2 and 2.1 eV in comparison with the free-standing graphene. We note that CO adsorption appears insensitive to the SVHS, which might be attributed to its weak oxidation nature[36-37]. These therefore highlight the effectiveness of the SVHS enabled chemical activity.

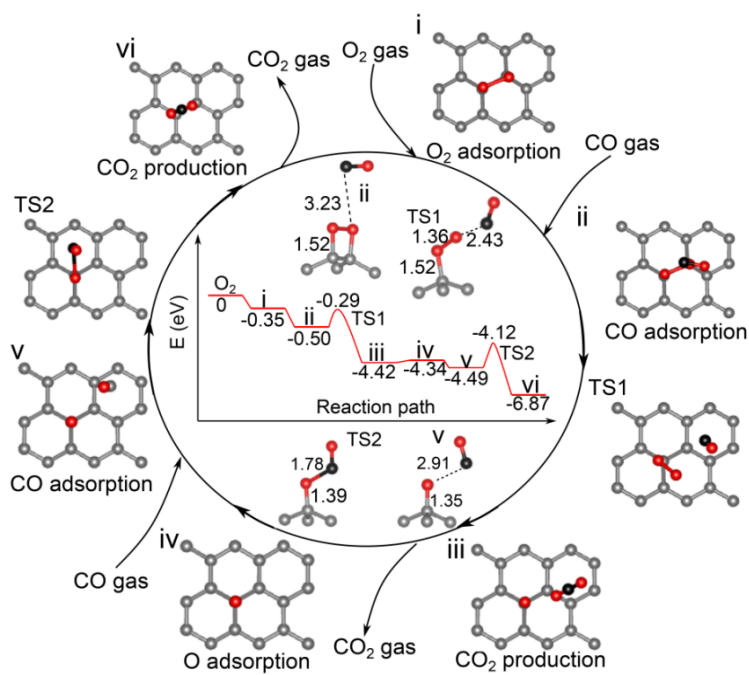

Figure 3: Catalytic cycle of CO oxidation on the Gra/Ca$_2$N. The transition states (TS) are marked, and the largest energy barrier is 0.37 eV for TS2. The insets in the cycle show the calculated energy profile and the configurations of the corresponding intermediates. The Ca$_2$N substrate is not shown for clarity.

It is well-established that there are two reaction mechanisms for CO oxidation by O$_2$, namely Langmuir-Hinshelwood (LH) and Eley-Rideal (ER) mechanisms [38-39]. The LH mechanism involves the co-adsorption of O$_2$ and CO molecules before reaction, while for the ER mechanism the free CO molecule directly reacts with the adsorbed O$_2$ molecule or O atom. Since the supported graphene only effectively bind O$_2$ molecule, the CO oxidation on the present heterojunction highly favors the ER mechanism. The reaction energy diagram along this pathway and the atomic configurations of the relevant intermediates are presented in Fig. 3. The incoming CO molecule directly attacks the pre-adsorbed O$_2$ species and is weakly adsorbed around O$_2$ molecule with the energy release of 0.15 eV [see Fig. 3(i-ii)]. Upon overcoming an ultra-low energy barrier of 0.21 eV (TS1), the first $CO_2$ molecule can be readily produced [see Fig. 3(iii)]. After desorption of the outcome $CO_2$ molecule, the remaining O atom further reacts with the incoming CO molecule to produce the second $CO_2$ molecule [see Fig. 3 (iv-vi)] with an energy barrier of 0.37 eV (TS2), which is the rate-determining step for the whole CO oxidation reaction. Such low energy barriers allow the reaction to occur far below the room temperature, suggesting the important role of SVHS in the CO oxidation reaction.

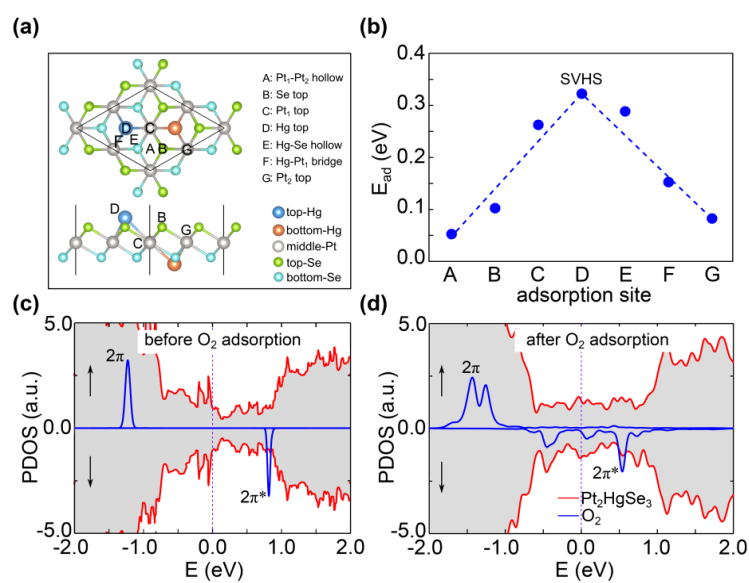

Figure 4: (a) Top and side views of the first layer of $Pt_2HgSe_3$ slab with the considered adsorption sites for $O_2$ from A to G. (b) The calculated adsorption energies of $O_2$ molecule on different adsorption sites of $Pt_2HgSe_3$ surface. (c) and (d) The PDOS of $O_2$ (blue line) and the first $Pt_2HgSe_3$ layer (red lines) before/after $O_2$ adsorption.

To further strengthen the viable role of SVHS in chemical reaction, we study the CO oxidation on $Pt_2HgSe_3$ (001) surface as another example. Fig. 4(a) shows the potential adsorption sites for $O_2$ and CO considered herein. It is found that only $O_2$ molecule is chemisorbed on the (001) surface by forming chemical bonds, similarly to the aforementioned heterojunction system. The $O_2$ adsorption energy is strongly site-dependent (Fig. 4(b)), and peaks at the top of Hg atoms, as expected, where the charge density of SVHS mainly localizes (Fig. 1(d)). The calculated adsorption energy is 0.32 eV at this optimal adsorption position, which corresponds to 0.42 e$^-$ transferring from the Pt$_2$HgSe$_3$ surface to fill the 2π* anti-bonding orbital of O$_2$ molecule [see Figs. 4(c) and 4(d)]. The CO oxidation reaction following ER pathway is also favored on the Pt$_2$HgSe$_3$ (001) surface with a low energy barrier of ~ 0.6 eV [see Fig. S3 in in Supporting Information]. Again, the SVHS in a distinct material system enables efficient CO oxidation.

In addition, we also studied the hydrogen evolution reaction (HER) on the Gra/Ca$_2$N heterojunction and Pt$_2$HgSe$_3$ (001) surface[40-41], for which the reaction activity is characterized by binding free energy of H ($\Delta G_H$), that is high HER activity corresponds to vanishing $\Delta G_H$ (in absolute value), and vice versa [41-42]. The results presented in Fig. S4 show that the free-standing graphene is extremely inert towards HER with $\Delta G_H$ ~ 1.7 eV [41], while, as expected, the SVHS in the supported graphene can significantly enhance the hydrogen binding strength (to ~ 0.1 eV) and the HER activity. Similarly, the topological Pt$_2$HgSe$_3$ (001) surface is also highly favorable with $\Delta G_H$ being as low as 0.16 eV at the Hg site [see Fig. S4(b)]. We hence have justified the viable role of SVHS in catalytic reactions, irrespectively of CO oxidation and HER.

Apart from the Gra/Ca$_2$N and Pt$_2$HgSe$_3$ (001) surface, other material systems with SVHS have been explored, such as Li/BN [18], alkali-metal intercalated graphene on SiC [17], twisted graphene bilayer [15-16], graphene/Y$_2$C electride heterojunction [see Fig. S5 in Supporting Information], Sn/p-type Si(111) [12] and Pb$_3$Bi/Ge(111) [13]. Given different orbital characteristics and its strong localization properties of SVHS, these systems are expected to possess distinct chemical activities toward different gas molecules (Fig. 2d), and be applied in different catalytic reactions. As an example, Li/BN with SVHS exhibits an ultrahigh reactivity towards the extreme inert $N_2$ molecule, with an adsorption energy of ~ 1.6 eV [see Fig. S2 in Supporting Information]. Such a platform may be capable for efficient ammonia synthesis, being motivated by recently discovered efficient conversion of $N_2$ to $NH_3$ using BN nanotubes filled with transition metal nanowires [43]. Additionally, we foresee that the catalytic performance of Gra/$Ca_2$N system constructed using twisted bilayer graphane instead of single graphene can be drastically enhanced considering the tunability nature of the position of VHS and the associated denser electronic states of twisted bilayer graphane reported in the recent experiments[44-46].

Finally, we compile the promising signatures of SVHS catalysis after having discussed its effectiveness on CO oxidation. The first one is the well-defined active sites uniformed distributing over the whole surface. Indeed, the charge density are highly localized around the catalytically active sites which are well-defined, as already shown in Figs. 1(c) and (d) for Gra/$Ca_2$N and $Pt_2HgSe_3$ (001) surface, respectively. This feature therefore makes SVHS catalysis outperform most of prevailing catalytic schemes with strongly localized active sites, such as heteroatom doping and atomic vacancy, which are distributed randomly and hard to identify[42, 47-48]. The other is the tunability nature of the catalytic activity with controlling the relative energy position of SVHS with respect to $E_F$. This can be inferred from the tunability of the $O_2$ binding strength and, simultaneously, the energy interval between SVHS and $E_F$, with the variation of interlayer distance in Gra/$Ca_2$N, as shown in Figs. 1(f) and 2(b).

To conclude, we demonstrate the SVHS catalysis through systematically studying CO oxidation reaction on the Gra/$Ca_2$N heterojunction and $Pt_2HgSe_3$ (001) surface. SVHS serving as an electron bath enables $O_2$ chemisorption and subsequently efficient CO oxidation, as justified by the low energy barriers of the rate-determining steps in the range of 0.2 ~ 0.6 eV for both material systems. The SVHS catalysis exhibits intriguing features of well-defined active sites, associated with the electronic states of SVHS, and tunablity nature with controlling the relative energy position of SVHS with respect to $E_F$. The present work enriches the Van Hove physics in material systems with reduced dimensions, and provides a new approach for the design of catalysts guided by the material electronic properties, beyond the d-band center theory.

## Acknowledgement

We thank Prof. Zhenyu Zhang at University of Science and Technology of China for helpful discussion. This work was supported by National Natural Science Foundation of China (Nos. 11774078, 11704005, and 11804077), China Postdoctoral Science Foundation (No. 2020M672201).